\documentclass[reprint,prb,twocolumn,showpacs,superscriptaddress,aps]{revtex4}

\usepackage{graphicx}
\usepackage{xcolor}
\usepackage{amsmath}

\newcommand{\redmark}[1] {{#1}}

\begin{document}

\author{R. Ortiz }
\affiliation{
 Departamento de F\'isica Aplicada, Universidad de Alicante,  03690 Spain \\
}

\author{ J. L. Lado }
\affiliation{
 International Iberian Nanotechnology Laboratory (INL),
Av. Mestre Jos\'e Veiga, 4715-330 Braga, Portugal
}

\author{ M. Melle-Franco}
\affiliation{
CICECO, Departamento de Qu\'imica, Universidade de Aveiro, 
3810-193 Aveiro, Portugal}

\author{J. Fern\'andez-Rossier } 
\affiliation{
 Departamento de F\'isica Aplicada, Universidad de Alicante,  03690 Spain \\
}
\affiliation{
 International Iberian Nanotechnology Laboratory (INL),
Av. Mestre Jos\'e Veiga, 4715-330 Braga, Portugal
}
\title{Engineering spin exchange in  non-bipartite graphene zigzag edges}




\begin{abstract}

The rules that govern spin exchange interaction in pristine
graphene nanostructures are
constrained by the bipartite character of the lattice,  so that the sign 
of the exchange is determined by  whether magnetic moments are on the same sublattice or else. 
The   synthesis  of   graphene
ribbons with perfect zigzag edges
and a fluoranthene group  with a pentagon ring,
a defect that
breaks the bipartite nature of the honeycomb lattice,
has been recently
demonstrated.
 Here we address how the electronic and spin properties of these
structures are modified by such
defects, both for indirect exchange interactions
as well as the emergent edge magnetism, studied  both with DFT and   mean field
Hubbard model calculations.  
 In all instances we  find that the local breakdown of the bipartite nature at
the defect   reverts the sign of  the otherwise ferromagnetic correlations
along the edge, introducing a locally antiferromagnetic intra-edge coupling
and,   for narrow ribbons, also revert the antiferromagnetic inter-edge
interactions that are normally found in pristine ribbons.    Our findings show
that  these pentagon defects are a resource that permits to engineer
the  spin  exchange interactions in  graphene based nano-structures.
 
\end{abstract}

\maketitle

A  central concept in the vast field of carbon based nanostructures is the 
 fact that  their electronic properties can change dramatically depending on their atomic structure.
Thus,  graphite, graphene, nanotubes and fullerenes all share the same atomic scale building blocks, carbon atoms with  $sp^2$ chemical bond,    yet their electronic properties are very different\cite{meunier2016}.   
Many remarkable electronic properties of graphene and other $sp^2$ nanostructures,    such as electron hole symmetry\cite{sutherland86,naumis07,pereira2008}, the existence of zero energy modes\cite{sutherland86,inui1994} and the rules that govern spin exchange interactions\cite{lieb89,fujita1996,JFR2007,saremi2007} derive from  the bipartite nature of the honeycomb lattice. 

A bipartite lattice can be  split  in two interpenetrating sublattices, $A$ and
$B$, such that first neighbors of $A$ sites are always $B$ sites, and vice
versa.    Whereas in 2D graphene the wave functions have the same weight on
both 
   sublattices,  in structures where there are more atoms of one type than the
other, such as zigzag edges,  there are zero modes whose wave function is 100
percent sublattice polarized\cite{sutherland86,JFR2007}.  These states play a
crucial role in our understanding of one of the most exciting   theory
predictions regarding graphene so far, namely, the existence of local moments
with ferromagnetic correlations in sub-lattice imbalanced graphene structures,
such as zigzag edges
   \cite{fujita1996,son2006,JFR2007,JFR2008,jung2009,yazyev2010}, 
      graphene functionalized with
hydrogen\cite{yazyev2007,palacios2008,yazyev2010} and a variety of planar
aromatic hydrocarbons \cite{morita2011,melle2015}.  Whereas a direct
experimental  local probe of the  magnetization is still missing,  indirect
experimental evidence in full agreement with DFT and model Hamiltonian
calculations\cite{tao2011, wang2015,ruffieux2016,gonzalez2016} supports the
existence of sublattice polarized states that most likely host unpaired
electrons.

  Interestingly, the chemical approach recently reported by Ruffieux et al
\cite{ruffieux2016,wang2015} has produced both ribbons with large sections of
pristine zigzag edges as well as edges decorated with a  fluoranthene group
(FG),
\footnote{
The FG is defined by analogy with the fluoranthene molecule which structurally
comprises one napthalene group and one benzene group joined together by a
pentagonal ring. In our case, the napthalene group is replaced by the graphene
ribbon, so that the pentagonal rings covalently link zig-zag graphene edges
with benzene rings.  (See figure 1a).
}
 as those shown in Fig. 1a and 3a, that break the bipartite character of the
lattice, on account of the presence of a pentagon  at the edge.  This naturally
leads to the question that  we address in this work: what is the fate of edge
states, and the spin interactions they produce, in the case of zigzag ribbons
decorated with non-bipartite intrusions.   
  Previous work had addressed the  magnetism of  
  an individual octagon-pentagon pair  in {\it bulk} graphene\cite{lopez2009}
(away from the edge),  the magnetic properties of  a line of
pentagon-pentagon-octagon (558) defects, as a grain boundary in
graphene\cite{alexandre2012},  but the  properties of the recently
found\cite{ruffieux2016}  zigzag ribbons decorated with FGs  have remained
unexplored. 
  To address the question,    we consider both {\em indirect exchange
interactions}  between some extrinsic spins, mediated by the electrons in the
FG decorated ribbons, as well as the edge magnetism that emerges,  according to
our Hubbard model, treated within the non-collinear mean field
approximation\cite{lado2014a,lado2014b}
   as well as calculations based on density functional theory (DFT).  
  
  Our main findings are the following. First,  the presence of these defects
locally depletes both the edge states and the edge magnetization,  but edge
magnetism with ferromagnetic correlations  persists  at the  pristine sections
as long as the FG groups in the edge are not too close to each other.
Second, and more important,  the
exchange interactions of two zigzag segments separated by a single FG are
antiferromagnetic, whereas  the face to face spin correlations can become
ferromagnetic,  both results at odds with the case of pristine
ribbons\cite{fujita1996,son2006,JFR2008}.  
   

\begin{figure}[t!]
 \centering
  \includegraphics[width=0.95\columnwidth]{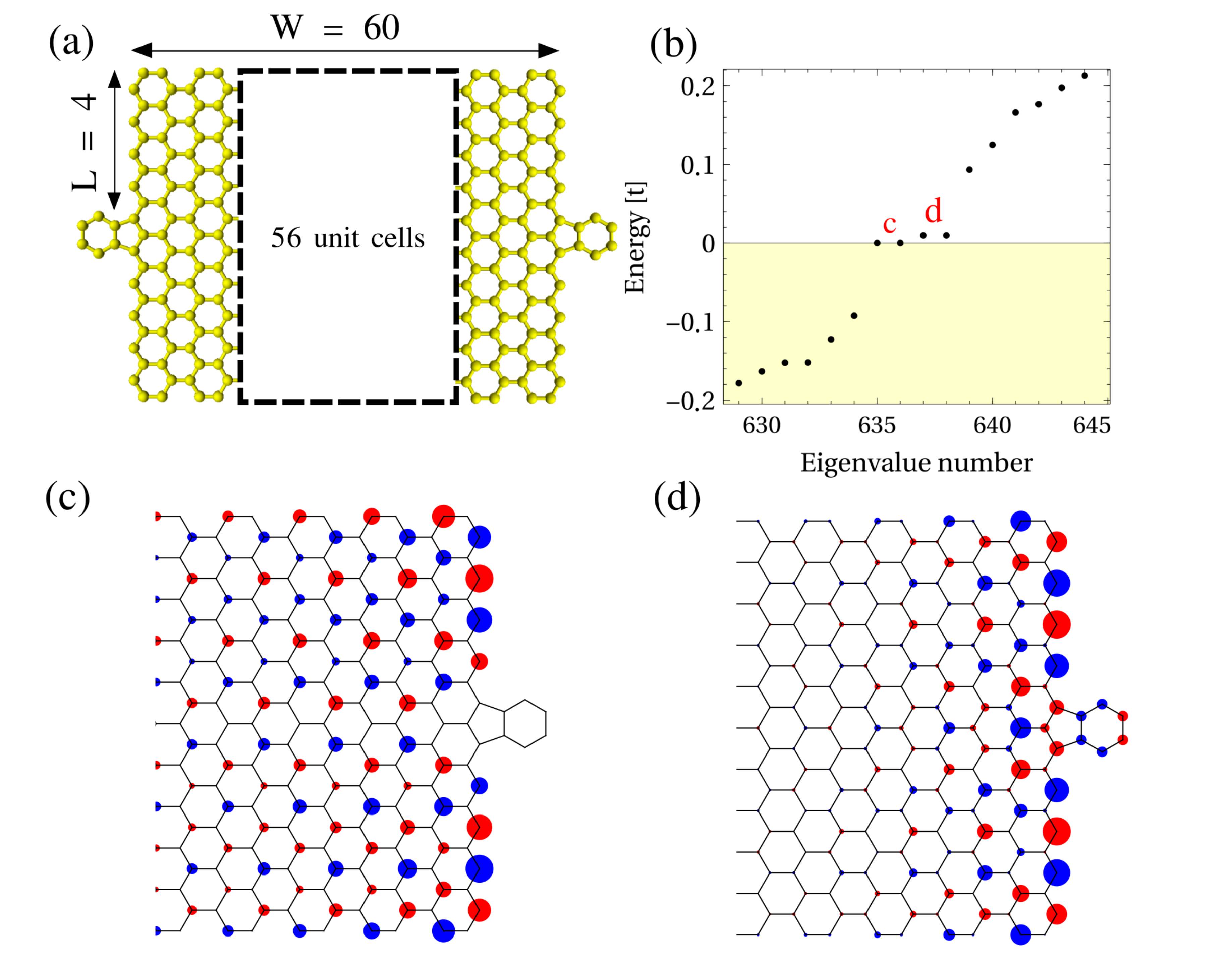}
\caption{$(a)$ Large graphene island with zigzag edges functionalized with a FG.
(b)  Energy spectrum, for eigenvalues close to the Fermi energy, showing four
in-gap states, two per right/left edge. The two fold
degeneracy in the smallest eigenvalues corresponds to the
two lateral edges.
Panels (c,d) are the wave functions labeled in (b),
showing the edge nature of the in-gap states, 
in particular their bonding and anti-bonding character.
Color in (c,d) labels the sign, whereas the size indicates
the magnitude of the component.
The splitting between the two states in the edge allows to map
the system into a two site tight binding model, which is expected
to develop antiferromagnetic correlations.
The width of the island was chosen so that the bulk is gapped.
\redmark{The dimensions of the island are shown in panel (a),
width of 60 unit cells and 4 carbon atoms per semi-edge.}}
\label{Structure}
\end{figure}
 In order to study the effect of the  edges decorated with FG, we consider
finite size graphene islands  (see Fig. 1).
 We first study them with  the standard tight-binding model with one orbital
and first neighbor hopping $t$.  As we show below, our DFT results indicate
that the presence of the FG   preserves the planar geometry of the system,   so
that the $\pi$ orbitals are still decoupled from the $\sigma$ orbitals,
validating this model.   For simplicity, we assume the same hopping integral
$t_{\text{C-C}}= -\text{t} =-2.7eV$ among all first neighbor
bonds, including also those of the defect.
In order to understand the properties of an individual edge,  we compute the
spectrum of  a   rectangular shape graphene island,  terminated with two short
zigzag edges with $L$ edge atoms each,  separated by a distance $W$.  
 We choose  $L=10$ hexagons  so that the bulk spectrum is gapped.   For the pristine ribbon  
 there are 3  in-gap    $E\simeq0$ states whose wave functions are localized at
each  edge,  and their wave function is sublattice polarized (not shown).
The addition of a single
FG  at each side  breaks the bipartite character of the lattice and has the following consequences. First,  
 the number of  localized in-gap  edge states is reduced  from 6 (3 per edge)
to 4 (2 per edge).   Second,   the remaining edge states are split in energy,
so that only 2 of them have $E\simeq 0$, the other two move upwards in energy.
This  breaks electron hole symmetry \footnote{For this reason, the choice of
sign of $t$ matters, in contrast to the case of bipartite systems}.   The
intra-edge splitting arises from the formation of a bonding non-bonding pair of
two modes that are   localized at both sides of the defect.
 In the experiment of Ruffieux {\em et al.}\cite{ruffieux2016} ), the distance
between zigzag edges was $W=$ 
 5 hexagons ,  so that hybridization between  states  at both edges 
takes place.

We now address the spin exchange properties of these edge states. We do that
using  two  complementary approaches:  the study of the so called indirect
exchange coupling\cite{brey2007,saremi2007,black-schaffer2010}, originally
proposed by  
Ruderman, Kittel, Kasuya, Yosida (RKKY) \cite{RKKY1,RKKY2,RKKY3},  and the
study of the emergent magnetism generated by Coulomb
repulsion\cite{fujita1996,son2006,JFR2007,JFR2008,yazyev2010}.  For the second,
we use two different  methods,  DFT calculations and mean field Hubbard model
calculations, that happen to give very similar results.   In the case of
pristine graphene, it is well known that both RKKY indirect exchange and the
spin interactions between  magnetic moments that spontaneously emerge due to
Coulomb interactions  
comply with the rule that  interactions between 
  magnetic moments on the same sublattice are  ferromagnetic,  whereas moments on opposite sublattices interact antiferrromagnetically
  \cite{fujita1996,son2006,brey2007,saremi2007,JFR2007,palacios2008,yazyev2010,black-schaffer2010}.     
  
\begin{figure}[t!]
 \centering
  \includegraphics[width=0.95\columnwidth]{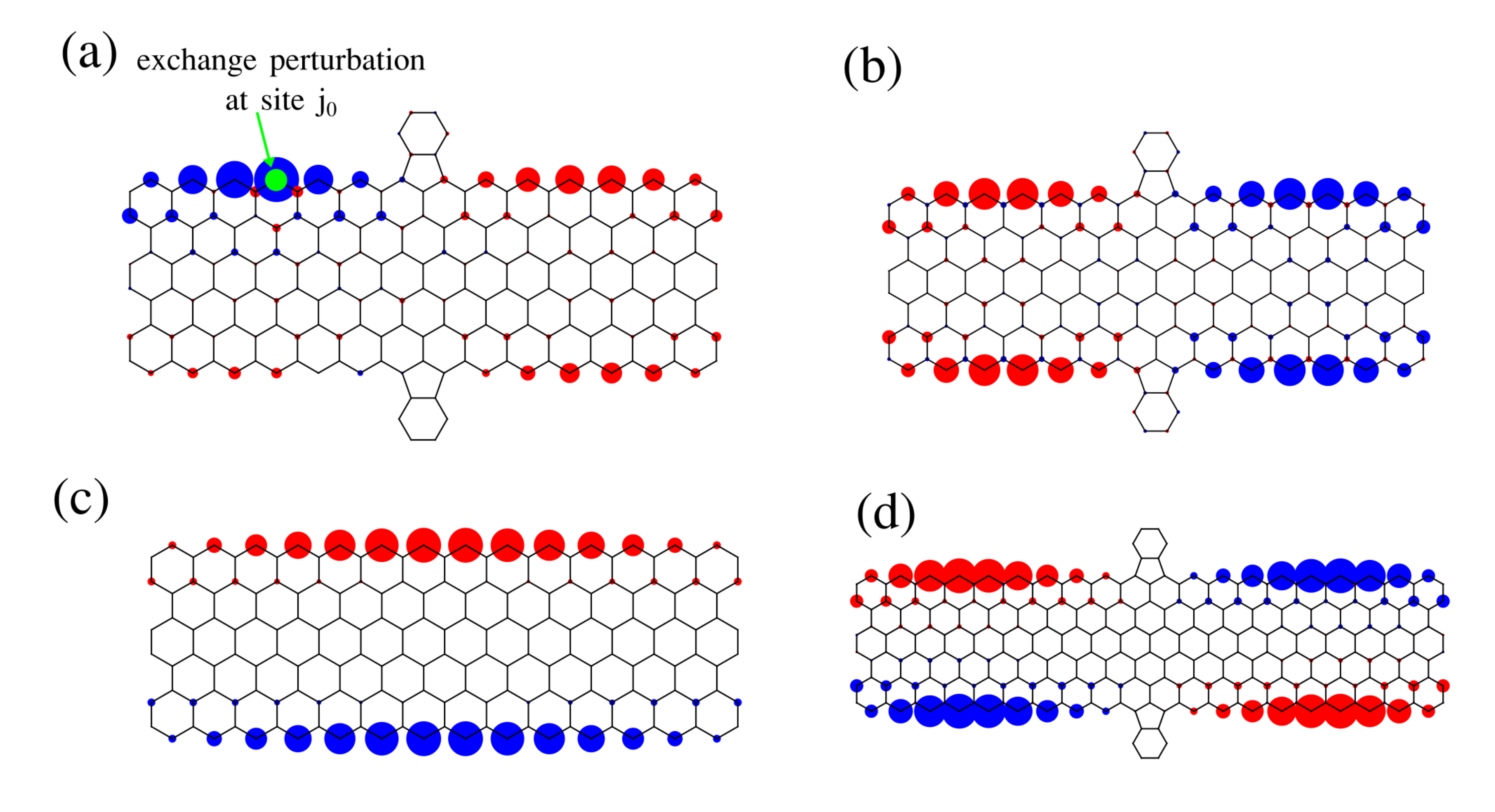}
\caption{$(a)$  Map of the non-local
spin susceptibility   $\chi_{ij_0}$, when $j_0$ is fixed 
at the green atom at the top-left edge,  and $i$ runs over the entire
island. The color reflects the sign of $\chi_{ij_0}$, 
and the size the magnitude. 
Panels (b,c,d) show the spatial
profile of the eigenstate $v$ of $\chi_{ij}$,
so that
$\chi_{ij}v_i = \lambda v_j$, with $\lambda$ the
largest eigenvalue, that characterizes the magnetic density right below the
critical temperature. The color represents the sign of the component
$v_i$, and the size its magnitude.
Panel (c) corresponds to the ribbon without FG, showing conventional
intra-edge ferromagnetic and inter-edge antiferromagnetic correlations.
Panels (b,d) show the antiferromagnetic
correlations in the same edge due to the FG group,
while the inter-edge correlations depend on the size of the island (b,d).
}
\label{RKKY}
\end{figure}

 The fact that  indirect exchange for spins on the same (different) sublattice
are correlated ferromagnetically (antiferromagnetically)   is a direct
consequence of the bipartite character of the
lattice\cite{lieb89,saremi2007,JFR2007}.  In order to see   how  this rule  is
modified due to the presence of the FG that  breaks the bipartite lattice  we
consider the Hamiltonian  ${\cal H}_0 + {\cal V}$, where ${\cal H}_0$ describes
the tight-binding model for the islands shown in figure \ref{RKKY}) and 
 ${\cal V}= J \sum_{\eta=1,2} \vec{m}_{\eta}\cdot\vec{S}_{\eta}$,
describes  two classical moments $\vec{m}_1$ and $\vec{m}_2$  that are
exchange coupled to the
local spin density $\vec{S}_i
 =\sum_{\sigma,\sigma'} \frac{1}{2} c^\dagger_{i\sigma} c_{i\sigma'}\vec{\tau}_{\sigma,\sigma'}$ of the 
 graphene electrons.  Here,  
  $\vec{\tau}_{\sigma,\sigma'}$ are the Pauli matrices.

  The indirect exchange interaction between the magnetic moments at sites $1$ and $2$ 
   is given by\cite{brey2007}
  \begin{equation}
{\cal J}_{12}= J^2   {\chi}_{12}\vec{m}_1\cdot\vec{m}_2
\end{equation}
where $ {\chi}_{ij}$ is the non-local   spin susceptibility function, 
defined as the variation of the spin density in site $i$ due to the application of a local Zeeman field $\vec{b}$ in site $j$: 
\begin{equation}
\langle \vec{S}_i\rangle = \chi_{ij}  \vec{b}_j
\end{equation}
The non-local spin susceptibility  can be obtained  analytically for 2D graphene  \cite{brey2007,saremi2007}, using linear response theory.  For the systems considered here this is not possible and, following  previous work\cite{black-schaffer2010},   we compute  the expectation value of the spin density  $\langle \vec{S}_i\rangle$ using the exact eigenstates
 of 
 ${\cal H}_0+  \vec{b}_1 \cdot\vec{S}_1
$, where  ${\cal H}_0$ is the tight-binding Hamiltonian of the island, and
$\vec{b}_1$ is a local magnetic field acting on atom $1$ only.   By doing so ,
we  obtain the entries   $\chi_{i1}$  of the susceptibility matrix.  Repeating
this procedure  changing the location of the perturbed site we obtain the
complete matrix.  For a system with $N$ atoms, this requires $N$
diagonalizations,  and for each of them , the computation of $N$ spin
densities. 

In figure \ref{RKKY}a we plot the map of $\chi_{ij_0}$, 
that represents the change
in
spin density induced in the sample when
we apply a local field in one of the edge atoms, labeled by $j_0$, for a
structure with one FG per zigzag edge, so that
there are four edge fragments.  
It is apparent that same fragment correlations are the largest
and ferromagnetic,  whereas interactions with the other fragments are
antiferromagnetic and smaller. These results are in contrast
with the indirect exchange interactions obtained for pristine graphene, that
are determined entirely by the sublattice degree of
freedom\cite{brey2007,saremi2007,black-schaffer2010}.  The fact that
the coupling of
one  edge fragment with all the others is antiferromagnetic   implies that
there will be some sort of  spin frustration.
\redmark{
This follows from the fact that
in a pristine ribbon, the correlation between opposite edges is
antiferromagnetic. Since the fluoranthene group also induces antiferromagnetic
correlations within the edge, the total system consists
on localized spins which are all of them correlated antiferromagnetically.
}
In the case of the figure, it
is worth noticing that the 
spin correlation with the moments located at the opposite edge are   weaker for
the moments located in front, compared with those in the diagonal.  This
anticipates the magnetic ground state that arises from the effective spin
Hamiltonian.

We now discuss the   inhomogeneous  magnetic order that would arise if we had a
set of classical magnetic moments $\vec m_i$ at each
site of the lattice, interacting with
the indirect exchange interaction
mediated by the  carriers in the graphene nano-island, governed by the
Hamiltonian 
$\mathcal{H} = \sum_{ij} {\cal J}_{i,j} \vec{m}_i\cdot\vec{m}_j$.  For that
matter we use the following method devised by P. W.
Anderson\cite{anderson1970}.
  We treat the magnetic moments in the mean field approximation, which permits
to write the free energy of the system as: 
\begin{equation}
{\cal F}(m_i) =\sum_i \left( \frac{\vec m_i^2}{2\chi_0} -    \vec{m}_i
\vec{b}_i\right)
\end{equation}
where $\vec{b}_i= \sum_j {\cal J}_{ij} \vec{m}_j$ is the effective field created by the interaction between the spins $m_i$, treated 
at the mean field level and $\chi_0=\frac{(g\mu_B)^2  S(S+1)}{3k_BT}$ is the
paramagnetic Curie susceptibility of the local moments.    In equilibrium, we
have $0=\frac{\delta {\cal F}}{\delta m_i}$ that leads to $\sum_j {\cal J}_{ij}
m_j = \frac{m_i}{\chi_0}$..  This equation is always satisfied by the
disordered non-magnetic solution, with $m_i=0$.    However it could also be
satisfied by the eigenvectors of the interaction matrix,  provided that their
eigenvalues $\lambda$ satisfy $\lambda= \chi_0^{-1}$. At very large temperature
$\chi^{-1}$ will be larger than the maximal $\lambda_{\rm max}$. However,  as
the temperature is reduced below $T_c$ such that $\lambda_{\rm
max}=\chi_0(T_c)^{-1}$ the system will order, and the magnetic order will be
given by the 
eigenstate $m^{\rm max}_i$ corresponding   to $\lambda_{\rm max}$. 

We apply this procedure to compute the magnetization map to  3 structures: a
pristine graphene ribbon \redmark{(figure \ref{RKKY}c)}, 
as well as two functionalized graphene ribbons with different lengths $L$
\redmark{(figure \ref{RKKY}b,d )}.   In all cases magnetism emerges at the
zigzag edge atoms, the bulk sites remain non-magnetic. In the pristine case, we
find that both zigzag edges would order ferromagnetically, with opposite
magnetizations,  expected from the standard RKKY in pristine graphene, and
given that all atoms in a given edge belong to the same sublattice.   In the
functionalized ribbons we also find 
ferromagnetic correlations between atoms on the same edge, provided that they
are not separated by the FG. The main novelty is the  antiferromagnetic
correlations between magnetic moments on the same edge, separated by  the FG.
The inter-edge correlations are  also different than in the pristine case   for
the shorter ribbon,  ferromagnetic, whereas in the longer one they are
antiferromagnetic.  The inter-edge correlations  reflect the competition
between the  antiferromagnetic couplings with a given edge fragment and the
other three.  

We now address the question of whether magnetic moments can arise in the functionalized zigzag  edges due to Coulomb interactions, as it was predicted to happen  in the case of pristine edges\cite{fujita1996,son2006,JFR2007,JFR2008,jung2009,yazyev2010}. There,   local moments with ferromagnetic correlations are expected on account of the  Hund's exchange between the degenerate  $E=0$ modes whose wave functions overlap along the edge.  The presence of the FG
functionalization  changes the situation (see figure 1),  giving rise to edge states that are linear combination of  states localized at both sides of the FG. 
In this scenario, Coulomb repulsion is expected to result in antiferromagnetic interactions between the edge portions separated by the defect, and  that is why  inter-edge correlations in pristine zigzag edges are antiferromagnetic as well. 

\begin{figure}[t!]
 \centering
  \includegraphics[width=0.9\columnwidth]{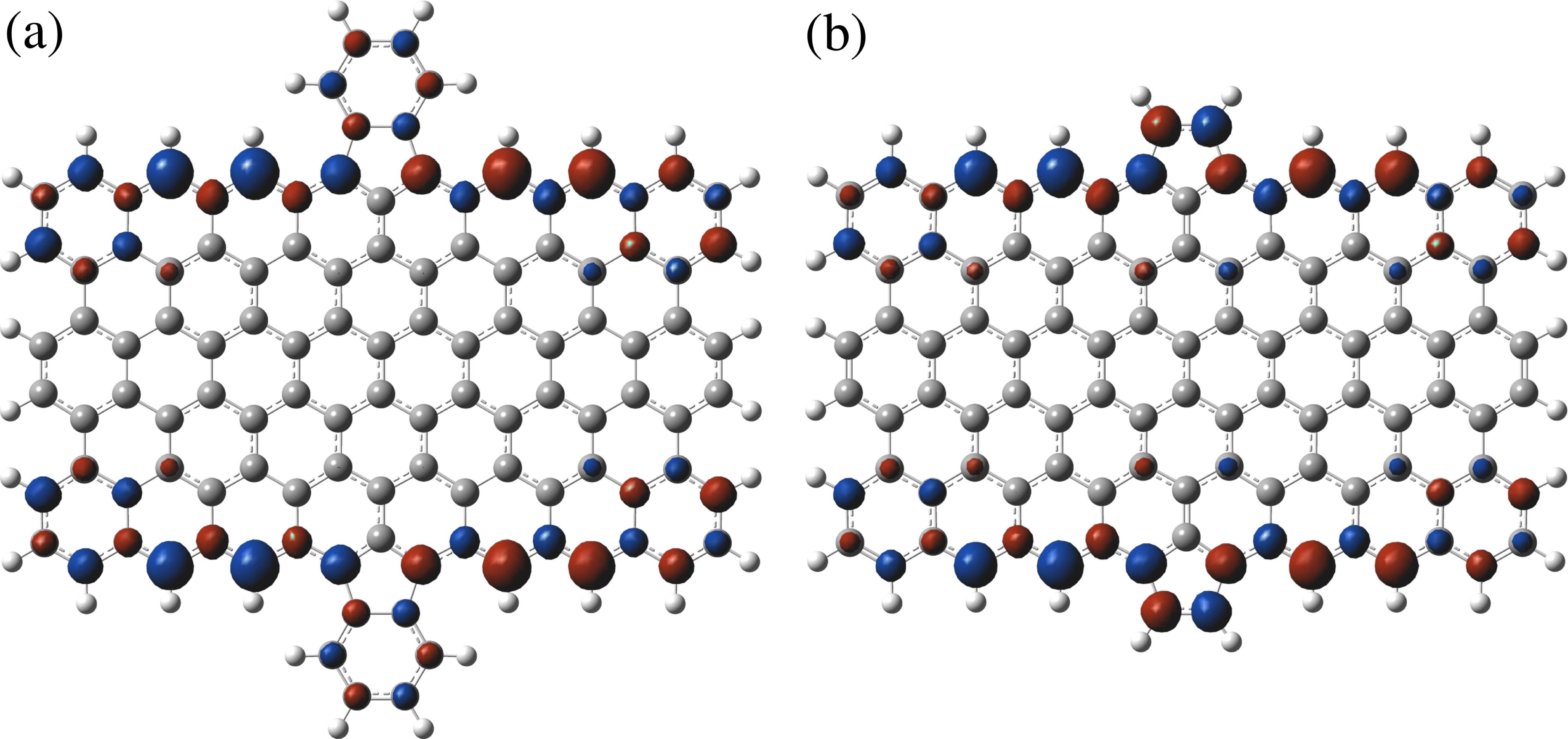}
\caption{ 
Iso-contours of the spin density obtained with DFT
calculations,
 for two different systems (a) and (b),
showing in both cases intra-edge antiferromagnetic
correlations. The color stands for the sign of the
magnetic moment. 
Calculations were performed with ${\rm CAM-B3LYP/6-31g(d,p)}$,
and the isosurfaces plotted correspond to isovalue 0.005.
  }
\label{DFT}
\end{figure}

 We discuss first the results obtained with DFT calculations.   We consider the
functionalized island of figure \ref{DFT}. The edge carbon atoms are passivated
with hydrogen. The calculation is done with the quantum chemistry code Gaussian
09\cite{g09}. The results shown here are obtained using the CAM-B3LYP
functional with the basis set 6-31g(d,p)\cite{melle2015}, yet similar results
were also obtained with B3LYP and PBE0 functionals. 

We computed three different spin configurations at the DFT level: zero spin,
antiferromagnetic and ferromagnetic. In all cases the molecular geometry was
relaxed until forces were below 0.024 eV/A and 0.013 eV/rad. Planar and
distorted initial molecular geometries yielded fully equivalent results upon
relaxation which is not surprising since both graphene and the fluoranthene
molecule are planar themselves. 
DFT yielded a magnetic ground state with all the functionals tried, namely:
PBE0, B3LYP and CAM-B3LYP. Unpolarized solutions appeared at higher energies
ranging from 0.3 (B3LYP) to 0.7 eV (CAM-B3LYP). Figure \ref{DFT}a depicts the
spin density for the antiferromagnetic state solution. The magnetization under
the FG  is depleted.  The arrangements of the magnetic moments
is  different from the one obtained for pristine edges, and in line with those
predicted by the RKKY interactions for this system.  We find an
antiferromagnetic coupling between the magnetic moments on the same edge that
are separated by the FG, whereas ferromagnetic
correlation between moments facing
each other in opposite edges.

The fact that this peculiar arrangement is originated by the breakdown of the
bipartite character of the lattice is confirmed by the results of the
calculation shown in figure \ref{DFT}b.  There we consider a functionalization
of a single pentagon, without the additional 4 carbon atoms that form the
external hexagon of the FG.
The magnetization profile obtained for this structure shares the same set of
spin correlations.  Therefore, it is the presence of the pentagon group the one
responsible of the antiferromagnetic coupling between same-edge atoms separated
by the defect.    Interestingly, a pentagon group like the one in in figure
\ref{DFT}b could be 
 formed by reconstruction of Klein ribbon edges\cite{klein}  or carbon nanotube
unzipping\cite{kleinPRB}. 
 

The emergent magnetism in zigzag edges can be described as well using the Hubbard model, given by
\begin{equation}
{\cal H}=-\sum_{\langle ij \rangle,\sigma} t c^\dagger_{i\alpha} c_{j\alpha}+
U \sum_i n_{i\uparrow}n_{i\downarrow}\equiv {\cal H}_0 + {\cal H}_{U}
\end{equation}
where $n_{i\uparrow}=c^{\dagger}_{i\uparrow}c_{i\uparrow}$ denotes the occupation operator of site $i$ with spin $\uparrow$ along an arbitrary quantization axis.  
 In  the mean field approximation
the results (magnetic moment density, energy spectrum) are very similar to those obtained with DFT\cite{JFR2007,JFR2008}.  The model has the advantage of being computationally less expensive and it can also be treated going beyond mean field interactions,  that permits to go beyond the broken symmetry analysis of magnetism and study thereby dynamic spin  fluctuations\cite{nuno2004}. 

Here we first treated the  model  at the non-collinear mean field
approximation\cite{lado2014a,lado2014b}. However, we always found collinear
solutions for this system, with all the moments along a common axis that we
take as the spin  quantization axis. With that in mind,
the
mean field  Hamiltonian
can be simplified into a collinear form:
\begin{equation}
{\cal H}^{MF}= {\cal H}_0 + U[\langle n_{i\uparrow }\rangle n_{i\downarrow }+
n_{i\uparrow }\langle n_{i\downarrow } \rangle
\rangle]
\label{mf}
\end{equation}
where  $\langle n_{i\uparrow}\rangle$ stand for the average of the occupation operator calculated within the ground state of the mean field Hamiltonian (\ref{mf}).
The mean field Hamiltonian is a functional of  $ \langle n_{i\sigma}\rangle$, which in turns depends on the eigenstates. 
This defines a self-consistent problem, that we solve by numerical iteration.
\redmark{In the following, unless stated otherwise, we take $U=t$.}

 \begin{figure}[t!]
 \centering
  \includegraphics[width=0.95\columnwidth]{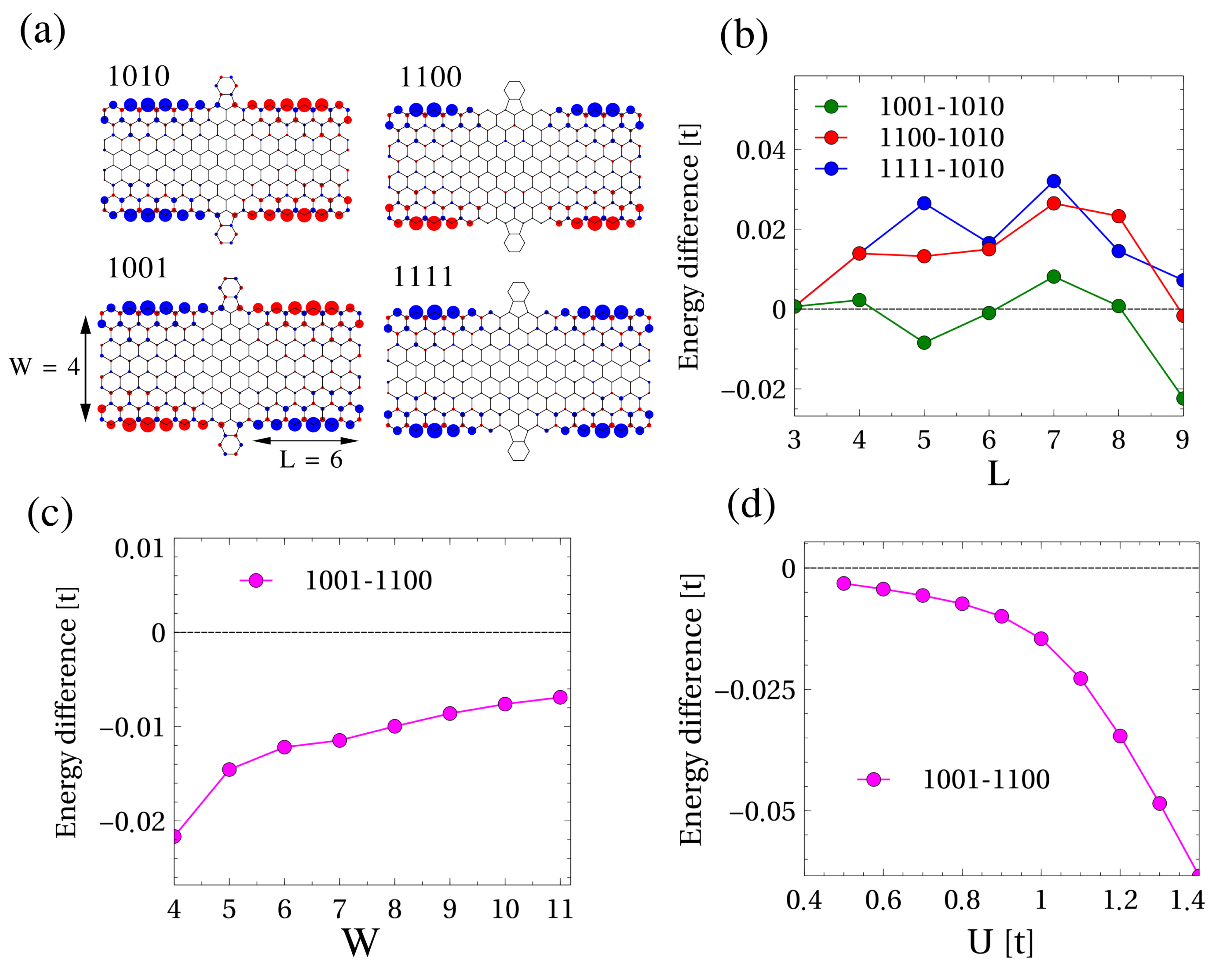}
\caption{ (a)  Magnetization profile of various self-consistent configurations 
 obtained within the mean field Hubbard model. Red and blue stand for the sign
of the magnetic moment. The size of the symbols stands for the magnitude of
the magnetic moment. 
(b)  Evolution of the energy difference between the 4 configurations shown in
panel (a) as a function of the 
size of the ribbon $L$. The lowest energy configuration are  always the $1010$
or the $1001$, the ones showing intra-edge antiferromagnetic correlations.
\redmark{
Panel (c) shows the energy difference between 1001 and 1100
as the island becomes thicker for $L=5$, whereas panel (d) shows the
energy difference as a function of U $(W=L=5)$.
Exponential extrapolation of
(c) to $W=\infty$ gives a total energy difference of 0.007 [t],
 19 meV for $t = 2.7$ eV}
}
\label{figure4}
\end{figure}

Adequate choice of the initial condition for the procedure can result in the
convergence of different solutions, that provide a local minima in the
landscape of different mean field solutions.   We obtain several of these
solutions for a series of graphene ribbons with one FG per edge,  shown in
figure  \ref{figure4} .   They all have spontaneous atomic magnetic moments at
the zigzag edges,  
$m_{i}=
\frac{\langle n_{i\uparrow }\rangle-\langle n_{i\downarrow}\rangle}{2}$
that order ferromagnetically in the edge fragments as long as they are not interrupted by a FG. 
The magnetization is depleted at the location of the defect.   The magnitude of
the magnetic moment away from the defects is the same than the one obtained for
pristine edges.

 The difference between the various  self-consistent solutions, shown in figure
\ref{figure4}a,  lies on 
  relative magnetization orientation of the four ferromagnetic fragments.  We
label the magnetic orientation  of a given edge fragment with respect to the
spin quantization axis  with either $0$ or $1$.   Using this notation, we can
label the  4 possible distinguishable magnetic states of the structures with
one defect per edge.  
    The mean field approximation  permits to compute their energies
     through the expression: 
   \begin{equation}
E_G= \sum_n \epsilon_n - U \sum_i \langle n_{i\uparrow}\rangle \langle n_{i\downarrow}\rangle 
\end{equation}
  where $\epsilon_n$ stands for the eigenstates of the mean field Hamiltonian and the  the sum runs over the occupied states only.
   We can compare the energies of the different magnetic configurations and
infer the effective spin couplings between the magnetized edge fragments.    In
figure \ref{figure4}b we show the evolution of the ground state energies as a
function of the    lateral dimension of the island, $L$.   It is apparent that
the dominant exchange coupling between different edge fragments is the
intra-edge antiferromagnetic coupling across a defect.  For sufficiently short
structures, the ground state  displays ferromagnetic correlations between edge
fragments located face to face. For structures with larger $L$  the ground
state has antiferromagnetic correlations between opposite edge fragments.
This, together with the analysis of the indirect exchange coupling, suggests
that the effective couplings between all fragments are antiferromagnetic.  For
shorter structures the diagonal coupling outweights the face to face
interaction and the situation is reverted for larger ribbons.
   The energy differences between different magnetic configurations is in the
range of $10^{-2}t\simeq 30$meV. \redmark{We also checked
the scaling of the intraedge exchange coupling as the
two edges become further apart (Fig. \ref{figure4}c),
obtaining an asymptotic value of the intra-edge exchange
of $9.5$ meV for graphene. The intra-edge AF configuration is the stable
in a wide range of $U$, as shown in Fig. \ref{figure4}d,
with the exchange coupling growing with $U$. }

\begin{figure}[t!]
 \centering
  \includegraphics[width=0.95\columnwidth]{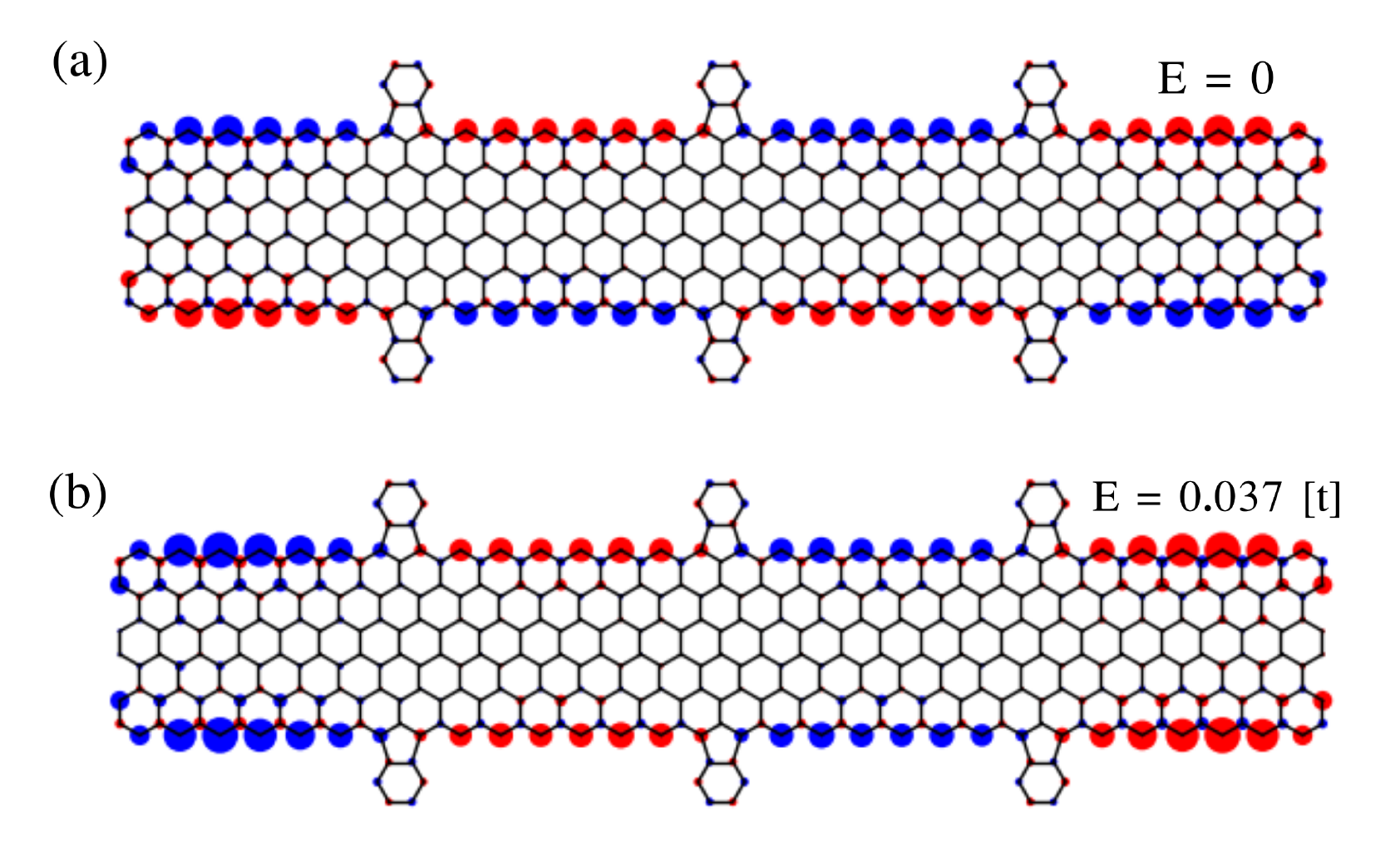}
\caption{$a)$  
Magnetization profile for  structure with several fluoranthene groups
obtained within the mean field Hubbard model $U=|t|$. 
Red and blue stand for the sign of the magnetic moment. The size of the symbols stands for the magnitude of  
$ m_i$. 
Panel (a) shows the ground state configuration,
whereas (b) an excited state.
}
\label{figure5}
\end{figure}

We now study how the magnetic properties of the edges evolve as we scale up
their size, keeping a similar density of defects.  We thus  consider longer
structures with several FGs at  the edge, in line with those reported by
Ruffieux et al\cite{ruffieux2016}.   In figure \ref{figure5} we show the
results of our mean field calculations for an  elongated ribbon with 3 defects
located symmetrically at both edges, 
 that define 4 zigzag edge fragments  per edge.  We find that the same rules that govern exchange in the presence of a single  FG,  still apply for larger structures, which is not entirely obvious given the delocalized nature of the edge states along the edge direction.     Therefore,  we conclude that the structures shown by Ruffieux {\em et al.}\cite{ruffieux2016} host magnetic moments localized  at the edges with antiferromagnetic correlations for moments separated by a FG group.

In summary,  we have demonstrated that   graphene ribbons with zigzag edges decorated with fluoranthene groups reported by 
 Ruffieux et al\cite{ruffieux2016}  host edge magnetic moments, very much like
their pristine counterparts, but the rules that govern the spin interactions
between different edge fragments are reversed compared to pristine edges.
Thus,  whereas in pristine ribbons 
magnetic moments on the same edge are always ferromagnetically correlated\cite{fujita1996,son2006,JFR2007,JFR2008,yazyev2010},    antiferromagnetic correlations are possible when 
moments in the same edge are  separated by a single FG.  The one to one
relation between the sign of the spin correlations and the  sublattice degree
of freedom no longer applies,  as the presence of a pentagon  breaks down the
bipartite character of the lattice.    Thus, rather than being an unwanted
defect on the ideal structures,  the fluoranthene groups can  used as a
resource to   nano-engineer  spin-based  quantum technologies   based on
nanographenes, and  could also    be used as platforms to study  spin liquid
physics, spin ladders,
Haldane chains and other interesting and non-trivial quantum magnetism phenomena.

{\em Acknowledgments}
We acknowledge financial support by Marie-Curie-ITN
607904 SPINOGRAPH.  
JFR acknowledges financial supported by MEC-Spain
(FIS2013-47328-C2-2-P) and Generalitat Valenciana
(ACOMP/2010/070), Prometeo.
This work is funded by ERDF funds through the Portuguese Operational Program for Competitiveness and InternationalizationÐ COMPETE 2020, and National Funds through FCT- The Portuguese Foundation for Science and Technology, under the project "PTDC/FIS-NAN/4662/2014" (016656),
CICECO-Aveiro Institute of Materials, POCI-01-0145-FEDER-007679 (FCT Ref. UID
/CTM /50011/2013) and ON2 (NORTE-07-0162-FEDER-000086)
J. L. Lado  thanks
the hospitality of the Departamento de Fisica Aplicada at
the Universidad de Alicante.


\bibliographystyle{unsrt}
\bibliography{biblio}{}

\end{document}